# Transferrable AlGaN/GaN HEMTs to Arbitrary Substrates via a Two-dimensional Boron Nitride Release Layer

Michael J. Motala,[1,2] Eric Blanton,[3] Al Hilton,[3] Eric Heller,[1] Chris Muratore,[1,4] Katherine Burzynski,[1,4] Jeff Brown,[3] Kelson Chabak,[5] Michael Durstock,[1] Michael Snure,[5]* Nicholas Glavin[1]*

[1]Materials and Manufacturing Directorate, Air Force Research Laboratory, Wright-Patterson AFB, OH 45433

[2]UES Inc., Beavercreek, OH 45432

[3]KBR, 2601 Mission Point Blvd Beavercreek OH 45431

[4]University of Dayton, Dayton, OH 45409

[5]Sensors Directorate, Air Force Research Laboratory, Wright-Patterson AFB, OH 45433

*Co-corresponding authors: michael.snure.1@us.af.mil, Nicholas.Glavin.1@us.af.mil

**ABSTRACT**

Mechanical transfer of high performing thin film devices onto arbitrary substrates represents an exciting opportunity to improve device performance, explore non-traditional manufacturing approaches, and paves the way for soft, conformal, and flexible electronics. Using a two-dimensional (2D) boron nitride (BN) release layer, we demonstrate the transfer of AlGaN/GaN high-electron mobility transistors (HEMTs) to arbitrary substrates through both direct van der Waals (vdW) bonding and with a polymer adhesive interlayer. No device degradation was observed due to the transfer process, and a significant reduction in device temperature (327 °C to 132 °C at 600 mW) was observed when directly bonded to a silicon carbide (SiC) wafer relative to the starting wafer. With the use of a benzocyclobutene (BCB) adhesion interlayer, devices were easily transferred and characterized on Kapton and ceramic films, representing an exciting opportunity for integration onto arbitrary substrates. Upon reduction of this polymer adhesive layer thickness, the AlGaN/GaN HEMTs transferred onto a BCB/SiC substrate resulted in comparable peak temperatures during operation at powers as high as 600 mW to the as-grown wafer, revealing that by optimizing interlayer characteristics such as thickness and thermal conductivity, transferrable devices on polymer layers can still improve performance outputs.

**INTRODUCTION**

High-electron mobility transistors (HEMTs) based on AlGaN/GaN are a promising technology for high-power radio frequency (RF) applications[1] due to their wide bandgap and the formation of a two-dimensional electron gas (2DEG) at the AlGaN/GaN interface.[2,3] These factors result in high breakdown fields (3 MV/cm), carrier densities (>$2\times10^{13}$ cm$^{-2}$), and electron velocities ($2.7 \times 10^7$ cm/s), which supports devices with high output power densities while operating beyond 100 GHz, greatly surpassing GaAs and Si technology. The exceptional RF performance of these devices has generated a demand for integration into diverse applications, circuits, platforms, and geometries, some of which are incompatible with GaN epitaxial growth and device processing (integration with Si and III-V devices, flexible polymer substrates, etc.), thereby driving the need for device transfer methods to integrate GaN HEMTs onto arbitrary substrates.

One significant limitation on the power density and performance of GaN HEMT devices is an insufficient rate of heat removal from the active device regions to the substrate and package. To overcome this limitation, GaN devices can be integrated with insulating high thermal conductivity substrates, like SiC, or diamond; however, these substrates are not always compatible with synthesis/fabrication conditions (e.g. insufficient lattice matching, low melting temperature). Transfer by van der Waals adhesion offers a strategy to improve the heat flow to the substrate without the need of an adhesion layer that would reduce heat flow to the substrate.[4,5] Prior studies on AlGaN/GaN HEMTs transferred from low thermal conductivity substrates to diamond using thin inorganic bonding layers have demonstrated lower device temperatures and improved DC and RF performance over devices on Si, but were still out-performed by devices on SiC.[6,7] This performance gap is due to a number of possible challenges including the high thermal resistance of the bonding layer reducing the impact of the high thermal conductivity of diamond, process induced stress, and device degradation.[8,9] Alternatively, AlGaN/GaN HEMTs grown on a BN/sapphire have been transferred to high-thermal conductivity Cu using metal bonding, demonstrating significantly lower device temperatures and improved DC performance due to efficient heat removal.[10] Trindade *et al.* also transferred GaN LEDs bound to diamond platelets by capillary bonding for improved electro-to-optical bandwidth.[11] Not all substrates lead to improved performance. For instance, the integration of GaN LEDs into hydrogels leads to a strong degradation of devices when operated at a comparable power as those on a foil.[12]

A promising approach for arbitrary transfer of high performing devices is to take advantage of the weak van der Waals (vdW) forces in two dimensional (2D) layered materials to enable a simple mechanical lift-off process.[13] This process of using 2D layers as a release layer was developed by Yasuyuki Kobayashi *et al.*,[14] and is referred to as Mechanical Transfer using a Release layer (MeTRE). MeTRE has been used for a wide range of materials and devices, process eliminates the use of chemicals, and does not damage device layers or induce surface roughness the way wet etching methods or laser lift-off (LLO) can.[15,16] The versatility of this approach to transfer both single high-performance devices[10,17] or a full wafer[18,19] enables integration onto a wide range of substrates. In this work, we demonstrate transfer of high performance GaN HEMTs produced on few-layer 2D BN to nearly any arbitrary substrate using both direct vdW bonding and thin adhesive bonding to meet integration needs of these devices on a variety of platforms.

**RESULTS AND DISCUSSION**

Ultra-thin BN layers allow for the subsequent growth of high-quality AlGaN/GaN HEMT structures by metal organic chemical vapor deposition (MOCVD).[20,21] The BN layer serves a dual-purpose in this role to both facilitate high-quality epitaxial growth of GaN HEMT structures as well as providing for a weak bonding structure for subsequent mechanical fracture, resulting in film lift-off. The basic epitaxial structure for the HEMTs consists of an initial 1.6 nm BN 2D buffer layer on sapphire followed by a 15 nm AlN nucleation layer, a 1.5 µm thick Fe-doped GaN buffer, 0.5 µm undoped GaN layer, 2 nm AlN insert layer, 17 nm $Al_{0.27}Ga_{0.73}N$ boundary layer, and 3 nm GaN cap, all grown by MOCVD and depicted schematically in Figure 1a. This structure was then used to produce the T-gated HEMTs shown in Figure 1b and c with 170 nm gate length and 50 µm width.

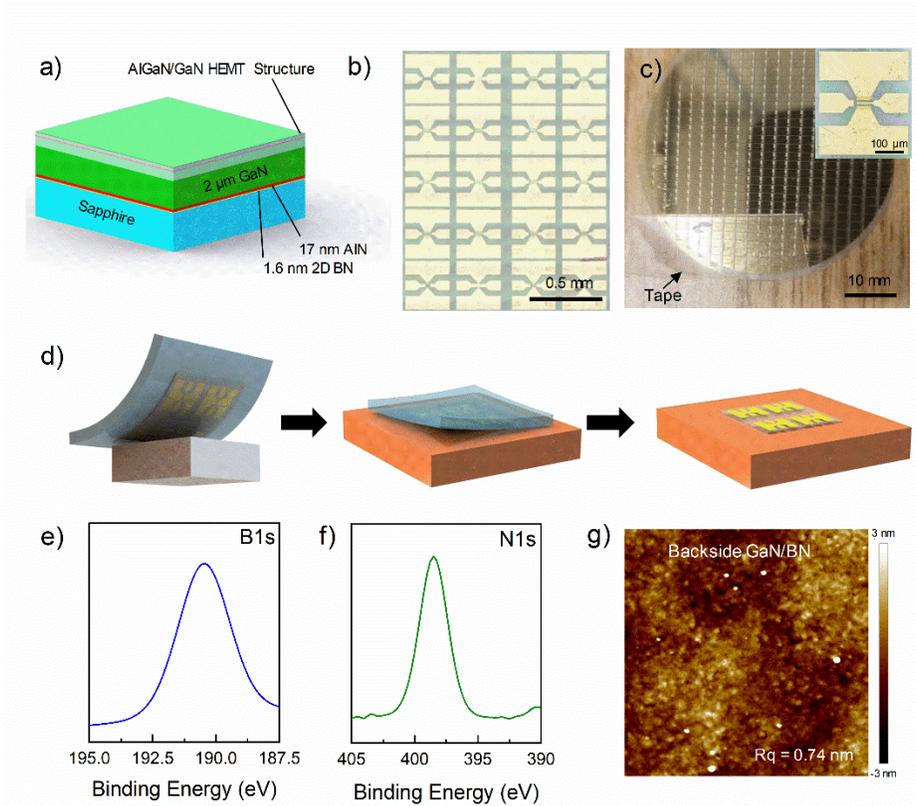

*Figure 1: GaN HEMT device layout and vdW lift-off. a) Material stack in GaN HEMT structure, b) Device layout, c) Image of wafer with water soluble tape for transfer; inset - close-up device image, d) schematic depiction of the transfer process, e) XPS spectrum of the B1s and f) N1s regions after transfer on backside of GaN, and f) AFM image of the same BN surface.*

The device transfer process flow is schematically depicted in Figure 1d. The lift-off of the GaN device layer is accomplished with the use of water soluble tape laminated to the surface followed by mechanical separation from the sapphire growth substrate. Attempts at removal via polydimethylsulfoxide (PDMS) stamping and quickly pulling off the surface via a kinetically controlled transfer process[22,23] were only partially successful as the interfacial bonding strength is fairly high, but some segments were removed in low yields. Conversely, the water soluble tape ensures a complete removal of the HEMT devices from the surface and is made possible because the tape adhesion strength to the device surface is greater than the vdW adhesion strength between the BN layers. The mechanical separation leaves a smooth BN layer underneath the GaN HEMT structure, as evidenced by the presence of BN on the backside GaN after removal from the growth substrate in the X-ray photoelectron spectroscopy results, and the smooth interfacial morphology, both shown in Figure 1e, f, and g respectively.

**Device Integration via van der Waals Transfer**

Following the device pick-up procedure, the device/tape stack was laminated to the semiconductor receiving surface without any adhesives in an attempt to bind the two materials together via strictly vdW forces. Lift-off and transfer occurred sequentially to ensure minimal ambient exposure, as vdW bonding between the receiving surface and back of the devices requires smooth interfaces that are free of surface contaminants. The devices were manually laminated to the receiving semiconductor substrate and the tape was removed by dissolution in warm water. After this tape removal step, the device surface was exposed to a low power oxygen plasma to remove any residual tape contaminants from the device surface. An optical image of a typical transferred device is shown in Figure 1b and the inset of Figure 1c. Cracking was observed due to strain relief upon pickup,[24] suggesting that minimizing the pickup area in subsequent transfers may reduce the incidence of cracking.

Two of the more common GaN substrates, sapphire and silicon carbide (SiC), were tested as receiving substrates for the direct transfer process. Figure 2 depicts the electronic performance of the vdW bonded AlGaN/GaN HEMT devices in the initial wafer configuration and direct bonded to sapphire and SiC substrates. Initial comparison of the output and transfer curves on sapphire before and after transfer in Figure 2a and b respectively show similar characters with little degradation caused by the transfer process. Both sets of devices exhibited a maximum transconductance ($G_M$) > 200 mS/mm and maximum source-drain current ($I_{DS}$) > 1.1 A/mm at $V_{DS}$ ~8 V. At high $V_{DS}$, devices on sapphire show a significant drop in $I_{DS}$, while at $V_G$ of +2 V, a ~ 10 mA decrease in $I_{DS}$ is measured. This nearly 20% drop in current is attributed to self-heating of the devices limited by the low thermal conductivity of sapphire (κ = 30 W/m·K). Only a slight drop in $G_M$ is observed in the transferred HEMTs back onto sapphire, presumably due to device-to-device variability or microcracks within the film upon transfer.[25] The performance of HEMTs vdW bonded to SiC in Figure 2c demonstrated an increase in $G_M$ to 250 mS/mm and reduction in the current drop from maximum current of only 4 mA (6%) at $V_G = 2$. These substantial improvements are presumed to be due to heat transfer from the device to the SiC substrate thermal conductivity, approximately ten times that of sapphire.[26,27] By comparison to the as-grown devices, the direct bonding process for transferred devices minimizes thermal interface conductance and demonstrates the utility of this transfer process for improving device performance.

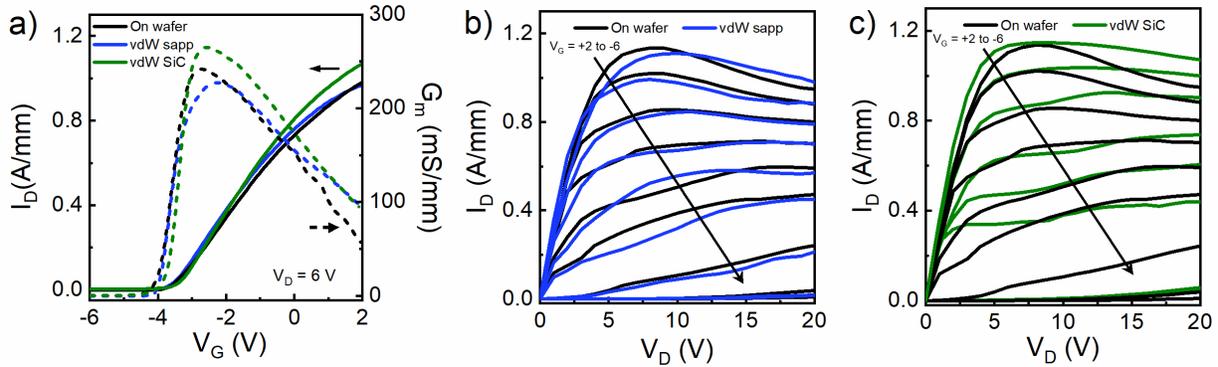

*Figure 2: Electronic performance of vdW-bonded GaN HEMTs. a) ID and $G_M$ sweep of on wafer and vdW-bonded GaN devices, b) IV family of GaN devices on Sapphire pre- and post-transfer, and b) IV family of GaN devices on wafer and vdW-bonded to SiC.*

**Device Integration via Thin Polymer Adhesive Layer**

If the receiving wafer is not smooth enough for vdW bonding (as in the case with a polymer or ceramic material), an adhesive is then necessary to facilitate integration. In this case, an adhesive layer of benzocyclobutene (BCB) was selected due to its high temperature stability, low outgassing, and capability to be spin coated in a thin layer. BCB also has a dielectric constant ranging from 2.6 to 2.8 that is relatively flat even in the high GHz range and at temperatures up to 200 °C.[28] Transfer of the GaN devices to a host substrate with an adhesive layer occurred first by spin coating the BCB (1.9 μm thick) onto the receiving surface and prebaking to drive off the solvent. Prebaking was necessary prior to binding otherwise the evaporating solvent can create bubbles.[29] The devices on the water soluble tape were then laminated onto the BCB surface and a thermal cure of the BCB was performed. Finally the tape was dissolved in warm water and the contact pads cleaned by plasma processing.

Figure 3a depicts the averaged peak drain metal temperature within the GaN HEMT device as measured with an IR camera as a function of applied power between the two source pads for the different substrates explored in this study. This is a common technique to understand thermal performance of the device as the metal layer is relatively uniformly heated and the GaN cannot be imaged accurately due to its low IR emissivity. In this depiction in Figure 3a, a lower slope indicates less device heating for the same operating power. The device performance on the original wafer is depicted by the black dots and serves as a baseline for comparison.

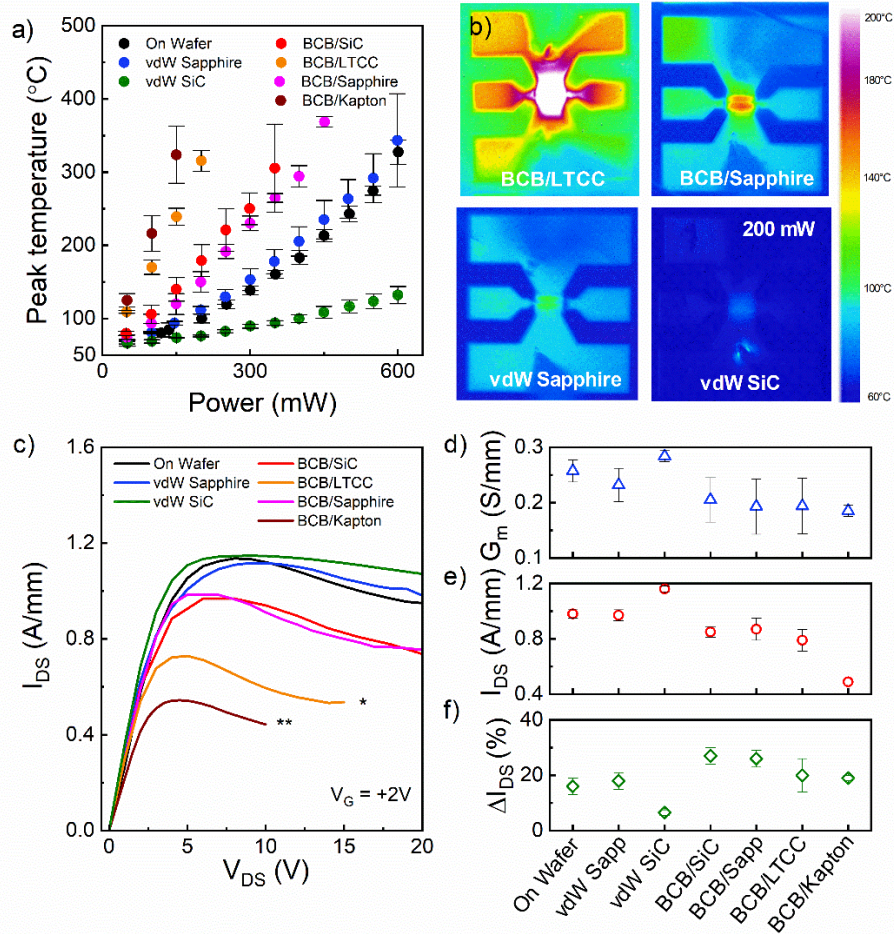

*Figure 3: Thermal performance of vdW and polymer adhesive bonded devices. a) Peak temperature observed in in-situ IR microscopy studies as a function of applied power, b) representative IR images of devices at 200 mW on various substrates, c) IV curves of AlGaN/GaN devices on substrates, d) peak conductance, e) max source-drain current, and f) percent current drop of HEMT devices.*

The transfer of the devices from the host sapphire wafer to an identical substrate did not result in a significant change in device operating temperature, as expected with the minimal change in IV behavior. All of the BCB-bonded devices (1.9 μm BCB thickness) showed a decreased power handling ability relative to the on wafer devices due to the lower thermal conductivity polymer interlayer (~0.29 W/m·K). Still, reasonable device performance curves were evaluated on sapphire, SiC, a low temperature co-fired ceramic (LTCC – GreenTape™) material, and even Kapton, all with the use of a BCB interlayer. Additionally, the vdW-bonded SiC expectedly showed a lower degree of device heating, where at 600 mW power the average operating temperature of devices transferred to the SiC substrate is just over 130 °C, compared to 326 °C as the measured operating temperature of the devices transferred to sapphire at the same power.

The devices transferred to Kapton and the LTCC could not be operated beyond 200 mW due to excessive self-heating. Despite the limited operating range, the GaN devices continue to function as expected and may be integrated into non-traditional substrate materials and geometries provided, they are operated with a slightly reduced power handling load. The LTCC substrate, for instance, ($\kappa$ = 4 W/m·K) is of interest for electronics integration as it allows the integration of printed resistors in a multilayer structure, facilitating the creation of a high-density package with reduced parasitic effects and fewer problems with electromagnetic compatibility. [30] The substrates also have acceptable dielectric properties at high frequencies [31] and they can be integrated with microfluidic structures.[32]

A visual comparison of the differences in the heat dissipation based upon substrate selection can be seen in Figure 3b. In this figure, selected IR thermal microscope images from different samples are shown with all the images taken at a fixed power of 200 mW. The samples were heated to a temperature of 60 °C on the stage to aid in the calibration of the IR images. The power was applied across the two device channels with the probes positioned on the larger device gate pads and the voltage being stepped until the desired power was obtained. The sample that produced the most electrically-induced heating, as depicted by the highest metal temperature of the selected images, is that of devices on BCB/LTCC. Within the measurement range, the temperature has been saturated, producing a large white spot at the center of the device, but the previously demonstrated device temperature from Figure 3a is around 300 °C. A noticeable drop in temperature is seen when comparing to the BCB/sapphire substrate and the direct sapphire transfer due to the lower thermal conductivity interlayer. The sample with the least amount of heating at 200 mW is that of a direct transfer to SiC, where only a small amount of heating relative to the background temperature of 60 °C is observed.

To better elucidate the influence of the interlayer material, we investigated DC performance of HEMTs bonded to high (SiC), medium (sapphire) and low (Kapton, LTCC) thermal conductivity substrates using a fixed BCB thickness of 1.9 μm. Representative $I_{DS}$-$V_{DS}$ output curves at $V_G = +2$ V for HEMTs on a wide range of substrates are depicted in Figure 3c. The output performance of HEMTs on substrates with no adhesive depict a steady increase in current drop at high $V_{DS}$, as well as a decreased $G_M$ and max $I_{DS}$ for the lower thermal conductivity substrates. With the addition of the BCB interlayer, the HEMTs on sapphire and SiC bonded with a 1.9 μm thick BCB layer depicted very similar values for current droop, $G_M$ and $I_{DS}$, which were considerably lower than on the growth wafer. The compiled $G_M$, max $I_{DS}$, and percent current drop are shown in Figure 3d. These results suggest that the device's thermal performance was dominated by the BCB layer and was not as strongly influenced by the underlying substrate, at least at thicknesses approaching 2 μm. Considering substrates with thermal conductivity comparable to the BCB such as Kapton and LTCC, we observe the substrate thermal conductivity again begins to dominate, limiting $G_M$, max $I_{DS}$ and causing a significant current drop. The self-heating of devices transferred to these substrates caused them to fail catastrophically at $V_{DS} < 20$ V.[33,34]

The reduced device performance upon integration with a BCB layer is primarily due to the low thermal conductivity relative to GaN and the underlying substrate, and therefore, optimizing the properties of the adhesive will be crucial in future device integration schemes. To better understand the relationship between substrate thermal conductivity and device heating, thermal simulations were carried out using COMSOL Multi-physics SW at 200 mW operating power. The resultant cross-section thermal images are depicted in Figure 4a-d with varying substrate and adhesive thickness conditions, including GaN on Sapphire, as well as 0 μm, 0.3 μm, and 1 μm BCB adhesive layer on SiC. The model assumed that the heat is dissipated uniformly within the channel, to match the open channel bias applied experimentally.[35] Thermal conductivities of 0.29 for BCB, $150 \times (300/T)^{1.4}$ for GaN, $35 \times (373/T)^{1.25}$ for sapphire, and $387 \times (300/T)^{1.49}$ for SiC are assumed (in units of W/m·K and where T is temperature in Kelvins). Clearly, as expected, the BCB layer resulted in a temperature drop across the interface and a resultant increase in temperature at the device region. This maximum device temperature is shown in Figure 4e based on these simulated results at varying thicknesses, indicating that significant changes in thermal device performance were experienced, especially within the first 2 μm of BCB layer and even up to a relatively thick layer of 10 μm.

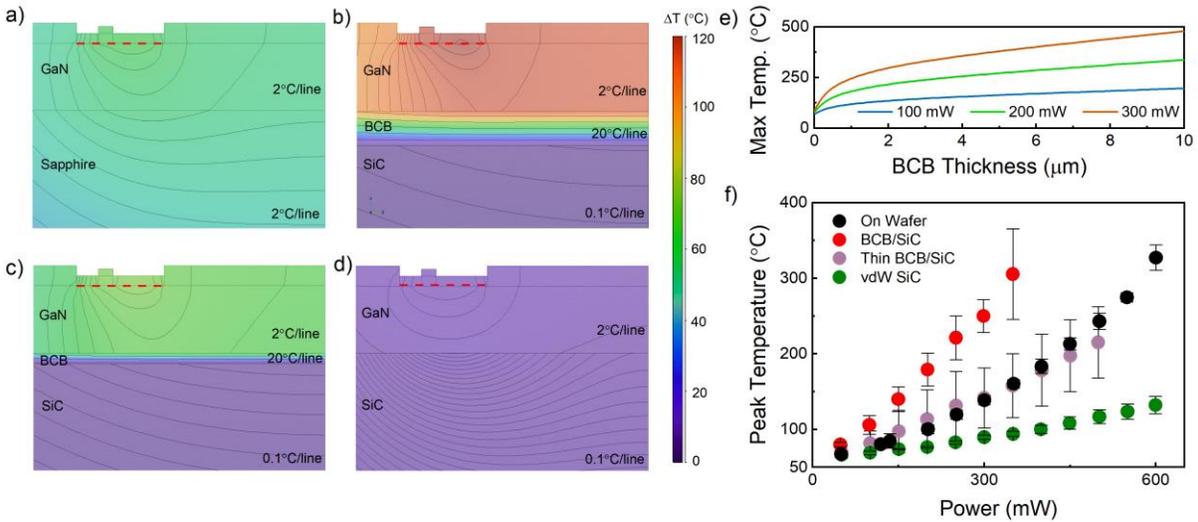

*Figure 4: Thermal device performance under varying BCB thicknesses. a) Cross-sectional thermal simulations of GaN/Sapphire, b) thermal simulations of 1 μm, c) 0.3 μm, and d) 0 μm BCB between GaN and SiC. In all cases, the cross-section through the material stack is shown at the middle of the finger, where temperatures are highest. All simulations are for 200 mW (2 W/mm) power dissipated uniformly across the channel region shown by the red dashed line, which is appropriate for open-channel bias condition and discussed in the methods section. Due to the huge variation in intrinsic material thermal conductivities, a hybrid contour plot is shown with 2 ° per division in both GaN and sapphire, 20 ° per division in BCB and 0.1 ° per division in SiC, designated as °C/line, e) simulated maximum device temperature as a function of power for devices with varying thicknesses of BCB, and f) IR imaging data comparing BCB thickness influence on peak temperature during device operation.*

To experimentally investigate the effects of the thickness of BCB, a thinner BCB layer was spun onto SiC substrates through dilution with mesitylene.[36-40] Upon reduction of the BCB layer thickness from 1.9 μm to 1.45 μm, a clear decrease in operating temperature was experimentally observed as depicted in Figure 4f. This is due to the increased heat transfer rate from the thinner BCB layer into the highly conductive SiC substrate, which has a measured thermal conductivity of ~ 370 W/mK at room temperature.[41] The devices operating at 300 mW exhibit an average temperature of 90, 141, and 250 °C for the vdW bonded SiC, thin BCB/SiC and thicker BCB/SiC, respectively. Even at 500 mW, the thinner BCB/SiC was able to be tested and depicted a maximum operating temperature of 215 °C compared to 116 °C for the direct SiC bonded device. Testing of this device was stopped at 500 mW in this case as operating much above 350 °C results in device failure during the measurement due to self-heating. The AlGaN/GaN HEMT device on a thin BCB/SiC substrate is still operating almost identically to the devices on the starting sapphire wafer, indicating that a precisely designed BCB thickness can retain thermal performance relative to the initial device configuration. Further reduction in device heating will be possible through either (or both) reduction in interlayer BCB thickness or improvement in adhesive thermal conductivity.

**CONCLUSIONS**

With increasing demand for high-frequency power amplification, AlGaN/GaN HEMT devices are becoming commonplace and the need to realize strategies that enable ubiquitous integration is clear. Herein we present integration of high performing GaN HEMT devices onto diverse substrates via either direct vdW bonding (if the substrate accommodates) or through a polymeric adhesive layer. The devices were initially epitaxially grown on a sapphire substrate with a 1.6 nm BN release layer on top of the sapphire, where the mechanical separation was shown to leave a smooth backside interface. Direct transfer without an adhesive layer was demonstrated onto sapphire while maintaining DC device performance and thermal properties. Integration onto a higher thermal conductivity SiC substrate showed improved thermal sink and a reduction in operating temperature by as much as 200 °C at 600 mW. On substrates that could not accommodate the direct bonding, a BCB interlayer was spun-cast and allowed for integration of GaN HEMT devices Kapton, a LTCC material, sapphire and SiC. Finally, by engineering this BCB layer to decrease the thickness, improved thermal performance of thin BCB/SiC substrates retained the starting performance on the as-grown sapphire wafer, indicating that further adhesive optimization can drastically improve device performance and reliability.

**EXPERIMENTAL METHODS**

**Device Integration.** The GaN HEMT devices were fabricated as described in the supporting information and in previous works.[17] After device fabrication, the GaN HEMTs were removed from the growth substrate manually. The process involved laminating 3M™ water soluble tape onto the surface, insuring that all air bubbles were removed, then peeling off the tape from the surface. This process is quite robust, offering high yields; and, due to the adhesion properties of the tape is a velocity independent, non-kinetic process.

For van der Waals transfer to semiconductor surfaces, the tape with the GaN HEMT devices was laminated to the surface and pressure applied. The semiconductor surface was free of any particulate matter that may inhibit contact between the backside of the device and the receiving surface. After sealing the surfaces together the sample was immersed into a warm water bath until the tape was fully dissolved.

Processing of the substrates with BCB starting with spin-coating an adhesion promoter, AP3000, at 3000 rpm for 30 sec. Next the BCB layer (Cyclotene, 3033-25) was spun at 2500 rpm for 30 sec. The solvent within the BCB was removed by ramping up to 100 °C at a 2 °C/min rate with a 30 min soak then brought back to room temperature. Once at room temperature the sample on the water soluble tape was laminated to the surface and a weight applied to the surface to ensure good contact. The BCB was cured by ramping to 100 °C at 2 °C/min followed by a 1 °C/min ramp from 100 °C to 175 °C with a 30 min soak. Higher curing temperatures were not done with the water soluble tape because oxidation of the tape made it difficult to remove. Once the BCB was brought back to room temperature, the tape was dissolved in warm water.

**Microscopy and Device Testing.** IR microscopy was preformed using a Quantum Focus Instruments (QFI) IR microscope with 15x magnification. Devices were loaded onto a temperature-controlled heat sink set to 60 °C. The devices were held in position by gravity, no thermal grease was used. The device operating temperatures were measured under a DC bias (between the larger gates pads) while manually stepping the power in roughly 50 mW increments. The thermal images are of the metal pads, which are an approximation of the temperature at the IR transparent GaN surface. The large 3 eV bandgap of GaN make it transparent to the infrared and visible radiation, which is known to lead to an underestimate of the surface temperature.[42] QFI images were averaged over 20 measurements.

**Thermal Simulations.** COMSOL Multiphysics finite element simulation software was used to solve the thermal diffusivity equation, with the heat load from electrical biasing assumed to be uniformly distributed within the rectangle defined by the 3.0 μm channel and the 50 μm width of the gate finger. This heat load was placed just under the passivation layer and on top of the GaN as seen by the red dashed line in Figure 4, which within about 20 nm precision is also where the resistive electron gas is located that is the source of the electrical heat load. This is a good approximation for the electrical biasing used, which puts the transistor in the linear or "fully open channel" state.[35] Reflection symmetry was assumed, with one finger modeled. The bottom of the substrate was assumed held at the ambient temperature (such as 60 °C) and all other surfaces were assumed adiabatic. The top surface was assumed to be plated with 600 nm Au (as per fabrication notes), following the shape seen in Figure 3a. Some other materials such as silicon nitride passivation on top of portions of this device structure, the ~20 nm thick AlGaN barrier layer, and some limited areas with Ohmic alloy were neglected in the simulation because poor thermal conductivity and thinness of layers precludes them from altering the thermal resistance appreciably.

AUTHOR INFORMATION

**Corresponding Authors**

*Michael.Snure.1@us.af.mil, Nicholas.Glavin.1@us.af.mil

**Author Contributions**

The manuscript was written through contributions of all authors. M.M., M.D., M.S. and N.G formulated the idea and drafted the manuscript. M.M., E. B., A. H., J. B., and K. C. performed device fabrication, IV characterization and IR camera testing, E.H., K.B., and C.M. worked on the and simulations and BCB characterization. All authors have given approval to the final version of the manuscript.

**Funding Sources**

This research was funded by the Air Force Office of Scientific Research under grant number FA9550-19RYCOR050.